\documentclass[aps,floatfix,nofootinbib,showpacs,espfig,twocolumn,amssymb]{revtex4}
\usepackage{graphicx}
\usepackage{latexsym}
\begin{document}
%
\title{UNIVERSAL  BEHAVIOR OF THE COEFFICIENTS OF THE CONTINUOUS EQUATION IN COMPETITIVE GROWTH MODELS}
\author{D. Muraca$^1$, L. A. Braunstein$^{1,2}$  and R. C. Buceta$^1$}

\affiliation{$^1$Departamento de F\'{\i}sica, Facultad de Ciencias Exactas y
Naturales\\ Universidad Nacional de Mar del Plata\\ Funes 3350,
$7600$ Mar del Plata, Argentina\\
$^2$Center for Polymer Studies and Department of Physics\\
Boston University, Boston, MA 02215, USA}

\begin{abstract}
The competitive growth models involving only one
kind of particles (CGM), are a mixture of two processes one with probability $p$ and the
other with probability $1-p$. The $p-$dependance produce crossovers between two different regimes.
We demonstrate that the coefficients of the continuous equation, describing their universality
classes, are quadratic in $p$ (or $1-p$). We show that the origin of such dependance is the
existence of two different average time rates. Thus, the quadratic $p-$dependance is a universal
behavior of all the CGM. We derive analytically the continuous equations for two CGM, in $1+1$
dimensions, from the microscopic rules using a regularization procedure. We propose
generalized scalings that reproduce the scaling behavior in each
regime. In order to verify the analytic results and the scalings,
we perform numerical integrations of the derived analytical equations.
The results are in excellent agreement with those of the
microscopic CGM presented here and with the proposed scalings.
\end{abstract}

\maketitle

Evolving growing interfaces or surfaces can be found in many
physical, chemical and biological processes. For example, in film growth either
by vapour deposition or chemical deposition \cite{family,barabasi}, bacterial growth \cite{albanobact} and
propagation of forest fire \cite{clar}.
The resulting interface has a rough surface that is characterized
through scaling of the interfacial width $W$ defined as
\begin{math}
W(L,t)=\left \{ \langle[h_i- \langle h_i \rangle]^2\rangle^{1/2} \right\}
\end{math},
where $h_i$ is the height at the position $i$, $\langle h_i \rangle = \sum_{i=1}^{L^d} h_i$ is the
spatial average, $L$ is the linear size, $d$ is the spatial dimension and $\left\{ \right\}$ denote configuration averages.
The general scaling relation \cite{family} for these growing interfaces  that evolves through a
single model can be summarized in the form $W(L,t) \sim L^\alpha f(t/L^z)$,
where the scaling function $f(u)$ behaves as $f(u) \sim u^\beta$
($\beta= z/\alpha$), for $u \ll 1$ and $f(u) \sim$ const for $u
\gg 1$. The exponent $\alpha$ describes the asymptotic behavior where the width saturates due to finite sizes
effects, while the exponent $\beta$ represent the early time
regime where finite-size effects are weak. The crossover time
between the two regimes is $t_s=L^{z}$.

The study of growth models involving one kind of particles in competitive processes (CGM) has received little attention, in spite of the fact that they are more realistic describing the growing in real materials, where usually there exist a competition between different growing processes. As an example, in a colony of bacteria growing on a substrate, a new bacteria can born near to another and stay there, move into another place looking for food or died. This ``bacteria'' can be thought as a particle undergoing either a deposition/evaporation process or deposition/surface relaxation.

The processes involved in the CGM could have different characteristic average time rate. Recently Shapir et  al. \cite{shapir} reported experimental results of surface roughening during cyclical electrodeposition dissolution of silver. Horowitz et al. \cite{albano} introduced a competitive growth model between random deposition with surface relaxation (RDSR) with probability $p$ and random deposition (RD) with probability $1-p$, called RDSR/RD. The authors proposed that the scaling behavior is characteristic of an Edward Wilkinson (EW) equation, where the coefficient associated to the surface tension $\nu$ depends on $p$. The dependance of $\nu$ on $p$ governs the transition from RDRS to RD. Using a dynamic scaling ansatz for the interface width
$W$ they found that the results are consistent provide that
$\nu \propto p^2$. Also Pellegrini and Jullien \cite{pellegrini1} have introduced CGM between ballistic deposition (BD) with probability $p$ and  RDSR with probability $1-p$, called BD/RDSR. For this model Chame and Aar$\tilde{a}$o Ries \cite{reis}
presented a more carefully analysis in $1+1$-$d$ and showed that
there exist a slow crossover from an EW to a
Kardar-Parisi-Zhang (KPZ) for any $p >0$. They also found that the
parameter $p$ is connected to the coefficient $\lambda$ of the
nonlinear term of the KPZ equation by $\lambda \sim p^{\gamma}$,
with $\gamma=2.1$.

In this letter, we show that the origin of such dependance is the
existence of two different average time rates. Thus, the quadratic $p-$dependance is an universal
feature of all the CGM. To our knowledge this is the first
time that the $p$-dependance on the coefficient of the continuous
equations is obtained analytical from the microscopic dynamics.

In order to test our hypothesis, we derive the analytical continuous equations for the local height for the RDSR/RD and BD/RDSR models. The procedure chosen here is based on regularization and
coarse-graining of the discrete Langevin equations obtained from a
Kramers-Moyal expansion of the master equation \cite{VK,Vvedensky,lidia}.

Let´s introduce first the general treatment of
this problem. Let us denote by $h_i(t)$ the height of the $i$-th
generic site at time $t$. The set $\{h_i,\,i=1,\dots ,L\}$ defines
the interface. Here we distinguish between two competitive
processes: A with probability p and average time of deposition $\tau_A$,
and B with probability 1-p and average time of deposition $\tau_B$.
In deposition processes with $p=1$ the average rate of deposition
is given by $\tau_0^{-1}=\left\{dh_i/dt\right\}_{(p=1)}$. If the
process is made with probability $p$ the average rate of deposition
is given by $\tau_A^{-1}=\left\{dh_i/dt\right\}_{(p)}=p\,\left\{dh_i/dt\right\}_{(p=1)}$.
The same hold for a process with probability $(1-p)$.
Thus, the particles are deposited at an average rate
\begin{equation}\label{Eq.tau}
\tau_A=\frac{\tau_0}{p}\;,\hspace{5ex}\
\tau_B =\frac{\tau_0}{1-p}\;.
\end{equation}
In the average time of each process, the height in
the site $i$ increases by
\begin{eqnarray} \label{Eq.rd}
h_i(t + \tau_A)&=& h_i(t) + a_{\perp}\,p\;R_i^A\;,\nonumber\\
h_i(t + \tau_B)&=& h_i(t) + a_{\perp}\,(1-p)\;R_i^B\;,
\end{eqnarray}
where $R_i^A$ and $R_i^B$ are the growing rules for
processes $A$ and $B$ respectively and $a_{\perp}$ is the vertical
lattice spacing. Expanding $h_i(t + \tau_A)$
and $h_i(t + \tau_B)$ to second order in Taylor series around
$\tau_A$ and $\tau_B$, we obtain
\begin{equation} \label{Eq.dt}
h_i(t + \tau_J)- h_i(t) \approx\frac{dh_i}{dt} \, \tau_J\;,
\end{equation}
for the process $J=A,B$. Thus, the evolution equation for the height (in the site $i$) for
this CGM is given by
\begin{equation}\label{Eq.3}
\frac{d h_i}{d t} =K_i^{(1,A)} + K_i^{(1,B)}+\eta_i(t)\;,
\end{equation}
where the first moments of the transition rate for each
process \cite{foot} are
\begin{eqnarray}\label{Eq.k}
K_i^{(1,A)}&=&\frac{a_{\perp}}{\tau_A}\,p\;R_i^A\;,\nonumber \\
K_i^{(1,B)}&=&\frac{a_{\perp}}{\tau_B}\,(1-p)\;R_i^B\;,
\end{eqnarray}
and the Gaussian thermal noise $\eta_i(t)$ has zero mean and covariance
\begin{equation}\label{Eq.4}
\left\{\eta_i(t) \eta_j(t')\right\} = a_{\perp}
\left(K^{(1,A)}_i + K^{(1,B)}_i\right)\delta_{ij}\,\delta(t-t')\;.
\end{equation}

In order to test our analytical result, we use two models. The
first model RDSR/RD considers a mixture \cite{albano} of RDSR
(process A) with probability $p$ and RD (process B) with
probability $1-p$. Let´s introduce the growth rule for each
process for the first model. In the RD growth model one chose a
column of a lattice, at random, among $L$ and a particle is
launched until it reaches the top of the selected column. The RDSR
is a variant of the RD: a particle is released from a random
position but when it reaches the top of the selected column is
allowed to relax to the nearest neighbor (nn) column if their
height are lower that the selected one. If the height of both of
the nn are lower than the selected one the relaxation takes place
with equal probability to one of them. For RD, $W(L,t)$ does not
depend on $L$, this means that the width $W$ does not saturate due
to the lack of lateral correlations. Thus, in this model: $W(t)
\sim t^{\beta_{RD}}$. Moreover, the RDSR model generates lateral
correlations, therefore one has $\beta_{RDSR}= 1/4$ and
$\alpha_{RDSR}= 1/2$. The first moment of the transition rate for
these processes are
\begin{eqnarray}\label{Eq.k1}
K_i^{(1,A)}&=& \frac{a_{\perp}}{\tau_A} \,p \left(
\omega_i^{(2)}+\omega_{i+1}^{(3)}+\omega_{i-1}^{(4)} \right)\;,\nonumber \\
K_i^{(1,B)}&=& \frac{a_{\perp}}{\tau_B} \,(1-p)\;\omega^{(1)}_i\;,
\end{eqnarray}
where the rules for both processes can we written as
\begin{eqnarray}\label{Eq.rules1}
\omega^{(1)}_i&=&1\;,\nonumber\\
\omega^{(2)}_i&=&\Theta(H^{i+1}_i)\;\Theta(H^{i-1}_i)\;,\\
\omega^{(3)}_i&=& \left\{\textstyle\frac{1}{2}\left[1-\Theta(H^{i+1}_{i})\right]+
\Theta(H^{i+1}_{i})\right\}\left[1-\Theta(H^{i-1}_{i})\right]\;,\nonumber \\
\omega^{(4)}_i&=& \left\{\textstyle\frac{1}{2}\left[1-\Theta(H^{i-1}_{i})\right]+
\Theta(H^{i-1}_{i})\right\}\left[1-\Theta(H^{i+1}_{i})\right]\;.
\nonumber
\end{eqnarray}
where $H^{i\pm s}_{i\pm k}=(h_{i\pm s}-h_{i\pm k})/a_{\perp}$,
and $\Theta(z)$ is the unit step function defined as
$\Theta(z)=1$ for $z\ge 0 $ and $\Theta(z)= 0$ for $z<0$.
The representation of the step function can be expanded as
$\Theta(z)=\sum_{k=0}^{\infty} c_k z^k$ providing that $z$ is
smooth. In any discrete model there is in principle an infinite
number of nonlinearities, but at long wavelengths the higher order
derivatives can be neglected using scaling arguments, since one
expect affine interfaces over a long range of scales, and then one
is usually concerned with the form of the relevant terms. Thus,
keeping the expansion of the step function to first order in his
argument and replacing the expansion Eq.~(\ref{Eq.rules1}), Eq.~(\ref{Eq.3}) can be
written as
\begin{equation}\label{Eq.dh-dt}
\frac{dh_i}{dt}= \frac{a_{\perp}(1- p)}{\tau_B} + \frac{a_{\perp}
p}{\tau_A} \left( 1 + c_1\,\frac{\Delta^2 h_i}{a_\perp}\right)+\eta_i(t)\;,
\end{equation}
where $\Delta^2 h_i= h_{i+1}- 2 h_i +h_{i+1}\simeq a_\parallel^2\;\partial^2 h/\partial x^2\rfloor_{h_i}$, and $a_{\parallel}$ is the horizontal lattice spacing. Replacing the rates
given by Eq.~(\ref{Eq.tau}) in Eq.~(\ref{Eq.dh-dt}) and
using a standard coarse-grain approach \cite{Vvedensky,lidia} the continuous
equation for this CGM is
\begin{equation}\label{Eq.c1}
\frac{dh}{dt}= F(p) + \nu(p)\,\frac{\partial^2 h}{\partial\,x^2}+
\eta(x,t)\;,
\end{equation}
where $h=h(x,t)$ and
\begin{eqnarray}\label{Eq.coef1}
F(p) &=& \frac{a_{\perp}}{\tau_0} \left[(1-p)^2 + p^2\right]\;,\\
\nu(p) &=&  2\;c_1\frac{a_\parallel^2}{\tau_0}\, p^2\;. \nonumber
\end{eqnarray}
The noise covariance is given by
\begin{equation}\label{Eq.cov}
\left\{\eta(x,t) \eta(x',t') \right\} = D(p)\;\delta(x -
x') \delta (t - t')\;,
\end{equation}
where
\begin{equation}\label{Eq.dif}
D(p)=a_\parallel\,a_\perp\, F(p)\;.
\end{equation}
Equations (\ref{Eq.coef1}) and Eq.~(\ref{Eq.dif}) shows that the quadratic dependance on
the coefficients of the continuous equation,  arises naturally
as a feature of the CGM and is due to the existence
of different average time rates.

The second model is a mixture of RDSR with probability $1-p$ ($B$
process) and ballistic deposition (BD) with probability $p$ ($A$
process) \cite{reis}. The evolution rules for RDSR are $\omega_i^{j}$, with $j=2,3,4$
[see Eq.~(\ref{Eq.rules1})]. In the BD model, the incident
particle follows a straight trajectory and sticks to the surface at the column $i$.
The height in the column $i$ is increased in
$\max[h_i+1,h_{i+1},h_{i-1}]$. If this process is done with
probability $p$ (A process), the rules can be summarized as:
\begin{eqnarray}\label{Eq.k2}
\omega^{(5)}_i&=&\Theta(H^i_{i+1})\; \Theta(H^i_{i-1})\;,\\
\omega^{(6)}_i&=&H^{i+1}_i\left[1-\Theta(H^i_{i+1})\right]\left[1-\Theta(H^{i-1}_{i+1})\right]\;,\nonumber\\
\omega^{(7)}_i&=& H^{i-1}_i\left[1-\Theta(H^i_{i-1})\right]\left[1-\Theta(H^{i+1}_{i-1})\right]\;,\nonumber\\
\omega^{(8)}_i&=&\textstyle\frac{1}{2}\; \delta(H^{i+1}_{i-1},0)\;
\left\{H^{i+1}_i\left[1-\Theta(H^i_{i+1})\right]\right.\nonumber\\
&&\hspace{14ex}+\left.H^{i-1}_i\left[1-\Theta(H^i_{i-1})\right]\right\}\;,\nonumber
\end{eqnarray}
where $\delta(z,0)=\Theta(z)+\Theta(-z)-1$ is the Kronecker delta.
Following the steps leading to Eq.~(\ref{Eq.c1}) the evolution
equation for this process can be written as:

\begin{equation}\label{Eq.c2}
\frac{dh}{dt}= F(p) + \nu(p) \;\frac{\partial^2 h}{\partial x^2}+
\lambda(p) \left(\frac{\partial h}{\partial x}\right)^2 +
\eta(x,t)
\end{equation}
where
\begin{eqnarray}\label{Eq.coef2}
F(p) &=& \frac{a_{\perp}}{\tau_0} \left[(1-p)^2 + c_0^2 \; p^2\right]\;,\nonumber\\
\nu(p) &=& \frac{a_\parallel^2}{\tau_0} \left[\textstyle\frac{1}{2}\,p^2 (1-c_0-2 c_0 c_1)+2 c_1 (1-p)^2\right],\\
\lambda(p)&=&\frac{a_\parallel^2}{\tau_0\,a_\perp}p^2\,c_1\,\left(5-4
c_0 -c_1\right)\;.\nonumber
\end{eqnarray}
The covariance of noise and $D(p)$ is given by Eq.~(\ref{Eq.cov}) and Eq.~(\ref{Eq.dif}), respectively.
Notice that we have to change $p$ by $1-p$ in
all the above equations for RDSR, because in the first model RDSR
is a kind $A$ process and now is a kind  B process.
Equation~(\ref{Eq.coef2}) shows again that quadratic dependance on
the coefficients of the continuous equation. The quadratic dependance
of $\lambda$ on $p$, found by Chame and Aar$\tilde{a}$o
Reis \cite{reis}, is a general feature of the CGM.

As both models have an EW behavior, it is expected that in that regime
the following generalized scaling ansatz \cite{natterman,albano},
\begin{equation}\label{Eq.scal}
W^2(p,L,t) \sim L^{2 \alpha}\;[D(p)/\nu(p)]\;
f\left(\nu(p)\,t/L^z\right)\;,
\end{equation}
where $f(u)\sim u^{2\beta }$ for  $u\ll 1$ and $f(u)\sim
\mbox{const}$ for $u \gg 1$. Moreover, the second model  is
represented by a mixture of  EW and KPZ universality classes. In
the early time regime  $W(t)\sim t^{\beta_{RDSR}}$, while a
crossover to a KPZ, with $\beta_{KPZ}= 1/3$ and
$\alpha\equiv\alpha_{KPZ}=1/2$, is expected in the intermediate
regime before the saturation. Thus, for the KPZ regime we propose
the following generalization \cite{amar} of the scaling behavior
of the width
\begin{equation}\label{Eq.scal-kpz}
W^2(p,L,t)\!\sim L^{2\alpha} [D(p)/\nu(p)]\,f\!\left(\lambda(p)
\sqrt{D(p)/\nu(p)}\,t/L^z\right)\,,
\end{equation}
where $z=3/2$, and $f(u) \sim u^{2
\beta_{KPZ} }$ for  $u \ll 1$ and $f(u) \sim\mbox{const}$ for $u
\gg 1$.

In order to test our analytical result and the proposed scalings, we
perform a numerical integration of Eq.~(\ref{Eq.c1}) and
Eq.~(\ref{Eq.c2}), and compute $W^2$ for both models.
\begin{figure}[h]
\includegraphics[width=5.5cm,height=5cm]{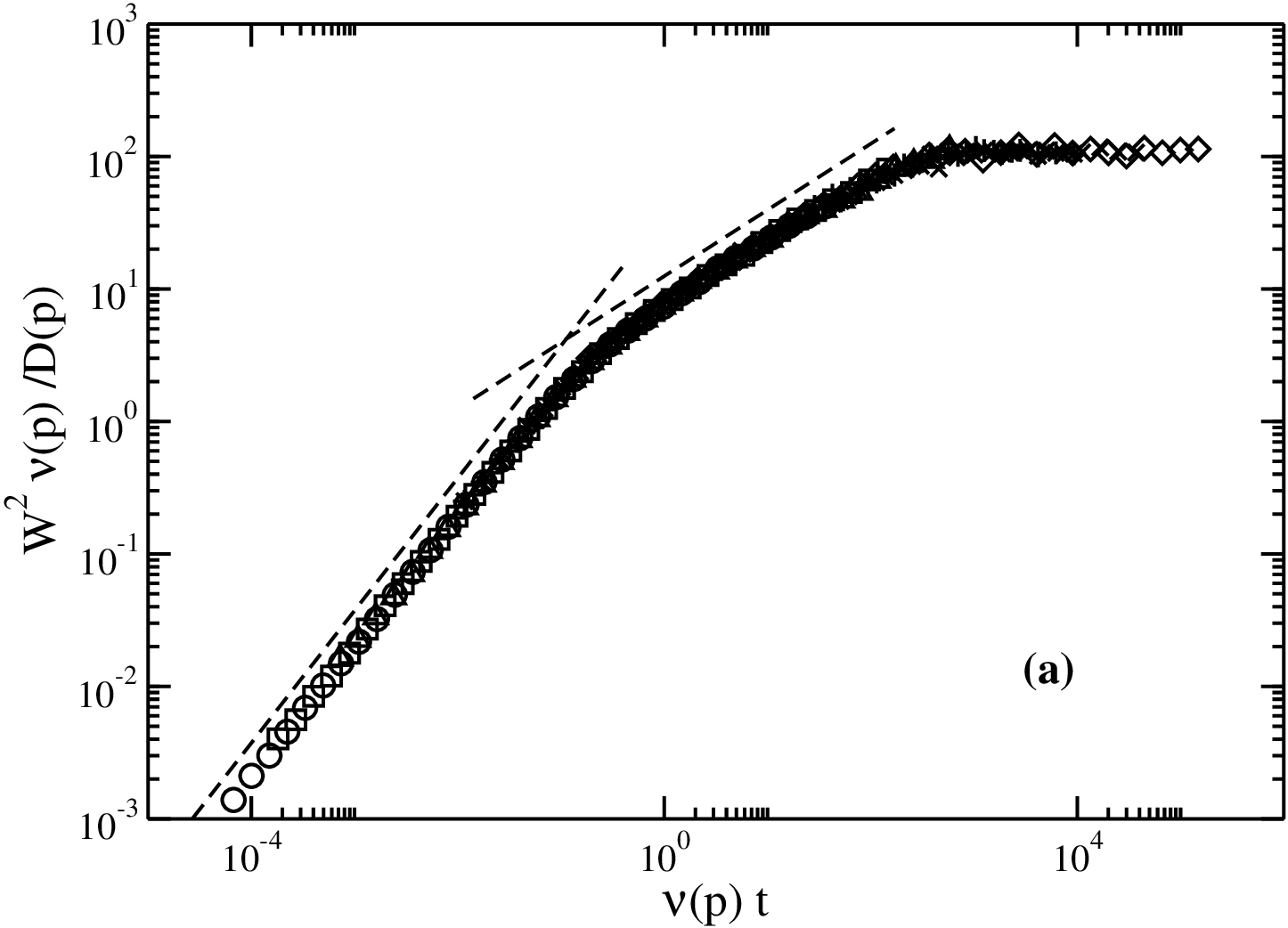}
\includegraphics[width=5.5cm,height=5cm]{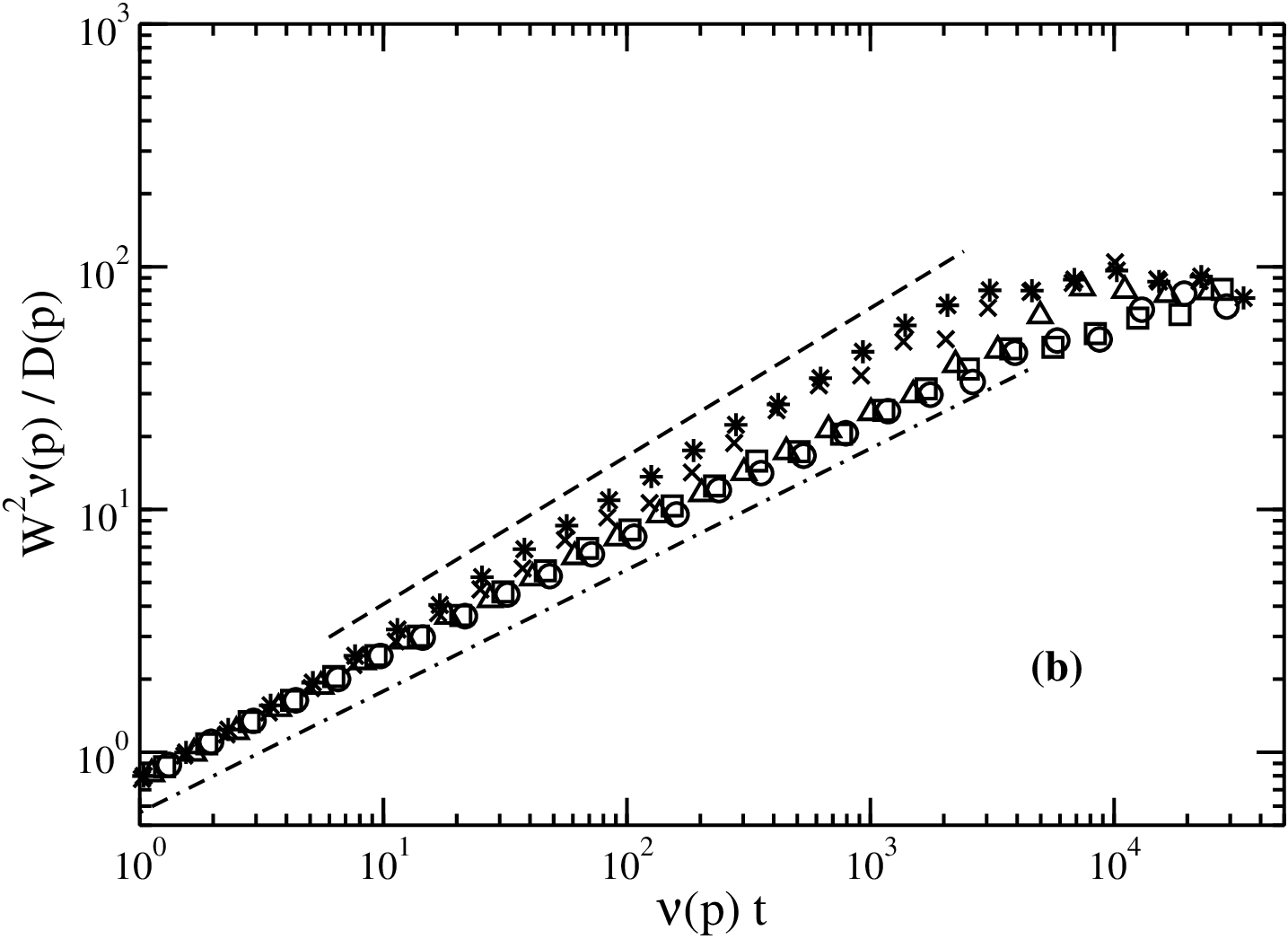}
\caption{
(a) Log-log plot of $W^2\nu(p)/D(p)$ for the RDSR/RD model as function of
$\nu(p)\,t$ for $L=128$. The different symbols represent different values of $p$,
$p=0.04$ ($\circ$), $p=0.08$ ($\Box$), $p=0.016$ ($\bigtriangleup$), $p=0.32$ ($+$), and
$p=0.64$ ($*$). Here we used $C=2.58$ and $b=0.2$ as parameters of the
$\Theta$-function representation. The dashed lines are used as guides to show the RD regime with $2 \beta = 1$ and
the EW regime with $2 \beta = 0.5$.
(b)  Log-log plot of $W^2 \nu(p) /D(p)$ for the BD/RDSR model as function of
$\nu(p)\,t$ for $L = 1024$. The symbols represent the same as in Fig.(1a).
Here we used $C=0.18$ and $b=0.5$. The collapse of the curves at the earlier stage clearly shows the EW
behavior ($2 \beta = 0.5$). After this stage the curves split and undergoes a slow
crossover to the KPZ behavior ($2 \beta = 0.66$). The dashed lines are used as guides to show the EW regime with $2 \beta = 1/2$ and the KPZ regime. The slope showed here is $2\beta=0.61$.\label{f1.1}}
\end{figure}
Notice that in order to numerically integrate the continuous equation, we
do need a continuum representation of the $\Theta$-function to
numerically compute the coefficients $c_0$ and $c_1$ related to
the ones of the continuous equations.
To perform the numerical integration, we chose the shifted
hyperbolic tangent \cite{Predota} as the continuous representation
of $\Theta$-function  defined as $\Theta(x)=\{1+ \tanh [C (x
+b)]\}/2$, where $b$ is the shift and $C$ is a parameter that
allows to recover the $\Theta$ in the limit $C \to \infty$.  The numerical integration
was made in short lattices using a discretized version of the
continuous equations Eq.(\ref{Eq.c1}) and Eq.~(\ref{Eq.c2}). The
results in large systems and the details of the integration are
beyond the scope of this letter and will be published elsewhere.

\begin{figure}
\includegraphics[width=6cm,height=5cm]{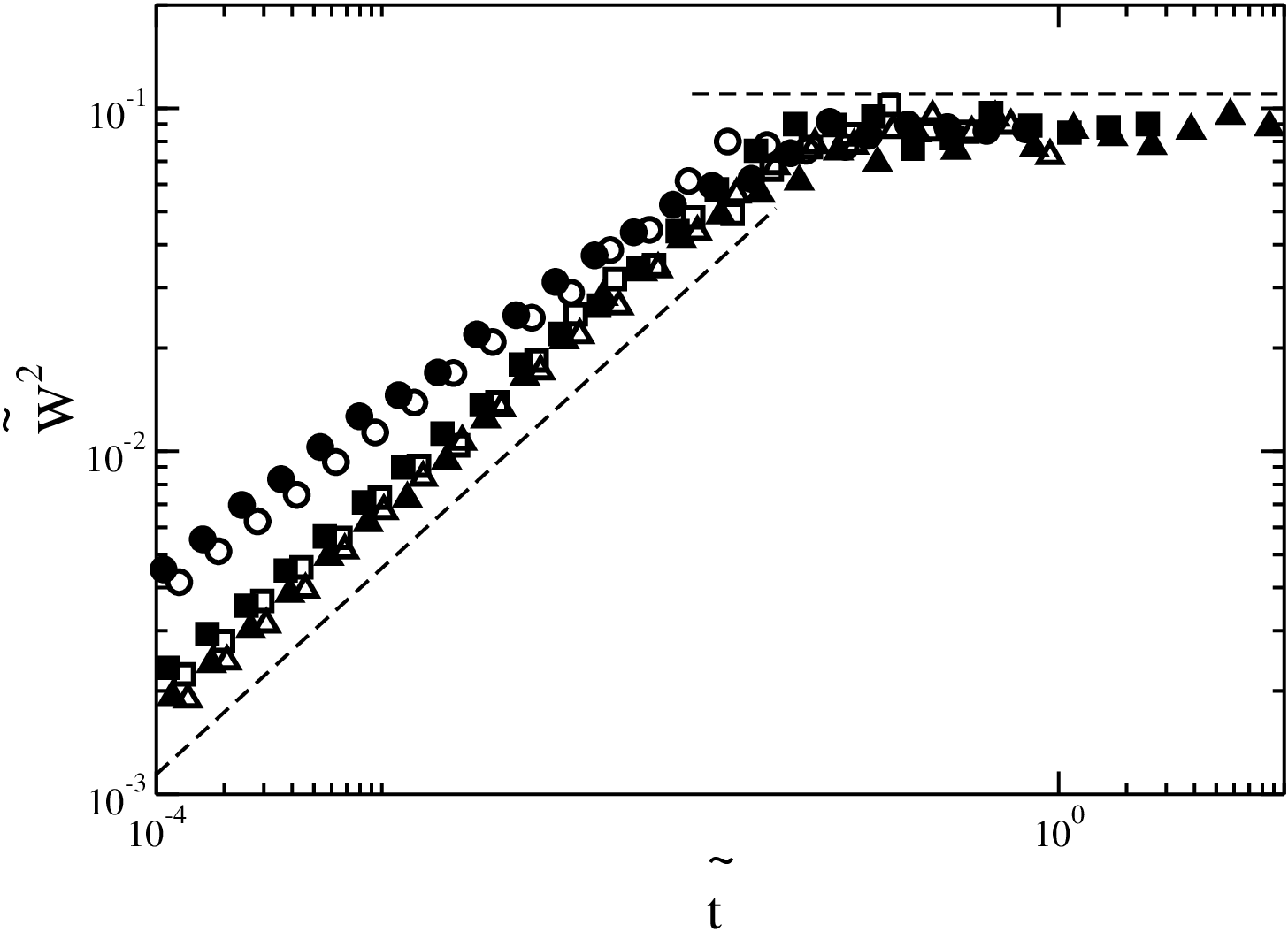}
\caption{Log-log plot of $\tilde{W}^2$ as function of $\tilde t$ as defined in the text, for $p=0.16$ ($\circ$), $p=0.32$ ($\Box$) and $p=0.64$ ($\bigtriangleup$). The empty symbols correspond to $L=512$ and the
filled ones to $L=1024$ . The collapse of the curves on the saturation regime using
$z=3/2$ shows that the curves saturates with a KPZ behavior as expected.\label{f1.2}}
\end{figure}

For the first model, Horowitz et al. \cite{albano} presented their
data from simulations plotting the scaling relation $W/L^\alpha
p^{-\delta}$ vs $t/L^z p^{-y}$. Clearly, their $\delta=1$ and
$y=2$ is related to our $\nu(p)$ and $D(p)$ [see
Eq.~(\ref{Eq.coef1}) and Eq.~(\ref{Eq.dif})]. In
Fig.~{\ref{f1.1}(a) we plot $W^2\,\nu(p)\,/D(p)$ as function of
$\nu(p)\,t$ for the that model for different values of $p$ and
$L=128$. This figure represent the same as in \cite{albano}
after coarse-graining. The agreement with the results of our
numerical integration, the numerical simulation \cite{albano} and
the scaling presented in Eq.~(\ref{Eq.scal}) is excellent. On the
other hand, for the second model, Chame and Aar$\tilde{a}$o Reis
\cite{reis} did not present the result for $W$. They studied the
crossover from EW to KPZ using an indirect method because of the
slow convergence of the discrete model to KPZ behavior. The crossover is
well represented in our Fig.~{\ref{f1.1}(b), where we plot the
same as in Fig.~{\ref{f1.1}(a) but for the second model. It is
clear the collapse of the curves in the EW regime. In the
intermediate regime the KPZ behavior appears thus, it is expected
that Eq.~(\ref{Eq.scal-kpz}) holds in that regime. In
Fig.~{\ref{f1.2} we plot
$\tilde{W}^2=W^2\,\nu(p)/[L^{2\alpha}D(p)]$ as function of
$\tilde{t}=\lambda(p)\sqrt{D(p)/\nu(p)}\;t/L^z$ for three
different values of $p$ using $z=3/2$. As $p$ increases, the KPZ
behavior appears earlier, but independent of $p$ all  the curves
saturate as a KPZ. The agreement with Eq.~(\ref{Eq.scal-kpz}) is
excellent in the saturation regime. The departure in the intermediate regime is due
to a slow crossover to the KPZ and to finite size effects.

Finally, notice that the quadratic dependence of the coefficients
of the continuous equation on $p$ is independent of the CGM
considered, because it is due to two different rates of deposition
given by Eq.~(\ref{Eq.tau}). This dependence is totally generally,
as shown from Eq.~(\ref{Eq.tau}) to Eq.~(\ref{Eq.k1}).

In summary, we demonstrate that the coefficient of the continuous equation have
quadratic dependance on $p$. This feature is universal for all the CGM model and
is due to the competition between different average time rate. We propose
generalized scaling for the model that reproduce the scaling behavior in each
regime. The numerical integration of the continuous equation are in excellent
agreement with the propose scalings and the numerical simulation of the models.

Acknowledgements: We thanks ANPCyT and UNMdP (PICT 2000/1-03-08974)
for the financial support.


\begin{thebibliography}{99}

\bibitem{family}F. Family, J. Phys. A: Math. Gen. {\bf 19}, L441 (1986).
\bibitem{barabasi}A.-L. Barab\'{a}si and H. E. Stanley, {\it Fractal Concepts in
Surface Growth}, Cambridge Univ. Press, New York (1995); P.Meakin, {\it Fractals, scaling and growth far from equilibrium}, Cambridge Univ. Press, Cambridge (1998).
\bibitem{albanobact} E. V. Albano, R. C. Salvarezza, L. V\'azquez and A. J. Arvia, Phys. Rev. B {\bf 59}, 7354 (1999).
\bibitem{clar} S. Clar, B. Drossel, and F. Schwabl, J. Phys.: Condens. Matter {\bf 8}, 6803 (1996).
\bibitem{shapir} Y. Shapir, S. Raychaudhuri, D. G. Foster, and J. Jorne, Phys. Rev. Lett. {\bf 84}, 3029 (2000).
\bibitem{albano} C. M. Horowitz, R. A. Monetti and E. V. Albano, Phys. Rev. E {\bf 63}, 066132 (2001).
\bibitem{pellegrini1} Y.P. Pellegrini and R. Jullien, Phys. Rev. Lett. {\bf 64}, 1745 (1990); {\it ibid.}
Phys. Rev. A  {\bf 43}, 920 (1991).
\bibitem{reis} A. Chame and F. D. A. Aar$\tilde{a}$o Reis , Phys. Rev. E  {\bf 66}, 051104 (2002).
\bibitem{VK} N. G. Van Kampen, {\it Stochastic Processes in Physics and
Chemistry}, North-Holland, Amsterdam (1981).
\bibitem{Vvedensky} D. D. Vvedensky, Phys. Rev. E {\bf 67}, 025102(R) (2003).
\bibitem{lidia} L. A. Braunstein, R. C. Buceta, C. D. Archubi and G. Costanza, Phys. Rev. E {\bf 62}, 3920 (2000).
\bibitem {foot}Notice that if the model involves only the nearest
neighbors the evolution equation contains only the first moment \protect\cite{Vvedensky2}.
\bibitem{Vvedensky2}D. D. Vvedensky, A. Zangwill, C. N. Luse, and M. R  Wilby, Phys Rev E {\bf 48}, 852 (1993); G. Costanza, Phys. Rev. E {\bf 55}, 6501 (1997).
\bibitem{natterman}T. Nattermann and Lei-Han Tang, Phys. Rev. A {\bf 45}, 7156 (1992).
\bibitem{amar} J. G. Amar and F. Family, Phys. Rev. A {\bf 45}, R3373 (1992).
\bibitem{Predota} M. P\v{r}edota and M. Kotrla, Phys. Rev. E {\bf 54}, 3933 (1996).
\end{thebibliography}
\end{document}